\documentclass[twocolumn]{article}
\usepackage[top=2cm, bottom=2.8cm, left=2cm, right=2cm]{geometry}

\usepackage{url}
\usepackage{graphicx}
\usepackage{amsmath}
\usepackage{amssymb}
\usepackage{xcolor}
\usepackage{float}
\usepackage{tabularx}
\usepackage{mathtools}
\usepackage[font=scriptsize]{subfig}
\usepackage[font=scriptsize,labelfont=bf]{caption}
\usepackage[super,numbers]{natbib}
\usepackage{sans}
\usepackage{authblk}
\providecommand{\keywords}[1]{\textbf{\textit{keywords---}} #1}
\usepackage{draftwatermark}
\SetWatermarkColor[gray]{0.7}
\SetWatermarkText{Accepted for Publication in \textsc{Medical \& Biological Engineering \& Computing} }
\SetWatermarkScale{0.9}

\begin{document}

\title{ Efficacy of the FDA nozzle benchmark and the lattice Boltzmann method
for the analysis of biomedical flows in transitional regime}

\markright{LBM simulations of transitional flow in the FDA nozzle}

\author[1]{Kartik Jain
  \thanks{Corresponding Author; E-mail: \texttt{k.jain@utwente.nl}}}

\affil[1]{
  \small Faculty of Engineering Technology,
  University of Twente, P.O. Box 217, 7500AE
  Enschede, \textsc{The Netherlands} 
}

\date{}

\maketitle

\begin{abstract}
  Flows through medical devices as well as in anatomical vessels despite being
  at moderate Reynolds number may exhibit transitional or even turbulent
  character.
  In order to validate numerical methods and codes used for biomedical flow
  computations, the U.S. food and drug administration (FDA) established an
  experimental benchmark, which was a pipe with gradual contraction and sudden
  expansion representing a nozzle.
  The experimental results for various Reynolds numbers ranging from $500$ to
  $6500$ were publicly released.
  Previous and recent computational investigations of flow in the FDA nozzle
  found limitations in various CFD approaches and some even questioned the
  adequacy of the benchmark itself.
  This communication reports the results of a lattice Boltzmann method (LBM)
  based direct numerical simulation (DNS) approach applied to the FDA nozzle
  benchmark for transitional cases of Reynolds numbers $2000$ and $3500$.
  The goal is to evaluate if a simple \emph{off the shelf} LBM would predict
  the experimental results without the use of complex models or synthetic
  turbulence at the inflow.
  LBM computations with various spatial and temporal resolutions are performed
  {\textendash}
  in the extremities of $44$ million to $2.8$ billion lattice cells 
  {\textendash}
  conducted respectively on $32$ CPU cores of a desktop to more than $300'000$
  cores of a modern supercomputer to explore and characterize miniscule flow
  details and quantify Kolmogorov scales.
  The LBM simulations transition to turbulence at a Reynolds number $2000$ like
  the FDA's experiments and acceptable agreement in \emph{jet breakdown
  locations, average velocity, shear stress} and \emph{pressure} is found for
  both the Reynolds numbers.


\end{abstract}

\section{Introduction}
  Computational fluid dynamics (CFD) has seen a growing interest in the
  biomedical community as flows within the cardiovascular system or in medical
  devices can be computed with sufficient ease to analyze various clinically
  and physiologically relevant details.
  However, the results from CFD can only be relied if the methods and software
  tools have been comprehensively verified and validated.
  The presence of turbulence in devices as well as in physiological flows poses
  additional challenges for the validation of CFD tools due to the complexity of the transitional flow physics itself.
  Furthermore, an appropriate quantification of such a flow regime requires fully resolved direct numerical simulation (DNS).

  The U.S. food and drug administration (FDA) established a benchmark in $1999$
  for this purpose~\citep{hariharan11a}.
  This benchmark was developed to contain features that could closely resemble
  those in medical devices including regions of flow contraction and expansion,
  flow recirculation, local high shear stresses as well as flow regimes ranging
  from laminar to transitional and turbulent.
  In addition, the benchmark was ensured to be simple enough such that CFD
  analyses could be readily performed and compared with experiments.
  The benchmark was thus a cylindrical pipe with a gradual contraction and sudden expansion.
  Particle image velocimetry (PIV) experiments were conducted by several
  laboratories on this device with Reynolds numbers $500$, $2000$, $3500$,
  $5000$ and $6500$ to study all the flow regimes namely laminar, transitional
  and fully developed turbulence.
  A large interlaboratory variability was found in experiments especially in
  cases with transitional flow regimes ($Re_{th} = 2000 \, \& \,
  3500$)~\citep{stewart12a}.
  The experimental datasets were publicly released with the intention of
  serving as a benchmark for the validation of CFD solvers thereof.

  \begin{table*}
    \centering
    \begin{tabular}{|l p{0.80cm} p{0.80cm} p{0.80cm} p{0.80cm} p{1cm} l p{2.0cm} r |}
      \hline
      Author group           & $Re_{th}$ $500$  & $Re_{th}$ $2000$  & $Re_{th}$ $3500$  & $Re_{th}$ $5000$  & $Re_{th}$ $6500$  & Numerical method     & Turbulence model & $Re_{crit}$   \\ \hline

      Present work           & $\Theta$       & \checkmark      & \checkmark        &$\Theta$         & $\Theta$        & LBM DNS              & none             & 2000        \\ 
      \citet{fehn19a}        & \checkmark     & \checkmark      & \checkmark        & \checkmark      & \checkmark      & high order DG        & none             & $\sim$ 2400 \\ 
      \citet{bergersen19a}   & $\Theta$       & $\Theta$        & \checkmark        & $\Theta$        & $\Theta$        & low order FEM        & none             & 3500        \\ 
      \citet{nicoud18a}      & \checkmark     & \checkmark      & \checkmark        & \checkmark      & $\Theta$        & fourth order FVM     & Sigma            & 3500       \\ 
      \citet{zmijanovic17a}  & \checkmark     & $\Theta$        & \checkmark        & $\Theta$        & $\Theta$        & fourth order FVM     & Sigma            & 3500        \\ 
      \citet{chabannes17a}   & \checkmark     & \checkmark      & \checkmark        & $\Theta$        & $\Theta$        & Taylor-Hood FEM      & none             & 2000        \\ 
      \citet{janiga14a}      & $\Theta$       & $\Theta$        & $\Theta$          & $\Theta$        & \checkmark      & low order FVM        & Smagorinsky      & $\Theta$  \\
      \citet{passerini13a}   & \checkmark     & \checkmark      & \checkmark        & $\Theta$        & $\Theta$        & low order FEM        & none             & 2000        \\ 
      \citet{delorme13a}     & \checkmark     & \checkmark      & \checkmark        & \checkmark      & $\Theta$        & high order FD, IBM   & Vreman           & 2000        \\ 
      \citet{white11a}       & \checkmark     &$\Theta$         &$\Theta$           &$\Theta$         & $\Theta$        & LBM                  & none             & $\Theta$  \\ \hline

    \end{tabular}
    \caption{Computational studies of flow in the FDA nozzle geometry in
    reverse chronological order.
    $Re_{th}$ refers to the Reynolds number at the throat of the nozzle.
    The signs \checkmark and $\Theta$ respectively refer whether or not a
    particular $Re_{th}$ was studied by the corresponding author group. 
    $Re_{crit}$ indicates the Reynolds number at which flow transitioned in the
    corresponding study.
    \textit{Part of the information is extracted from~\citet{fehn19a} } 
    }
    \label{tab:fdastud}
  \end{table*}

  Various numerical schemes ranging from finite differences to finite element
  and finite volume methods as well as high order discontinuous Galerkin (DG)
  methods have been employed by researchers to evaluate if their numerical
  method and the corresponding implementation can reproduce the results
  published by the FDA.
  Table~\ref{tab:fdastud} lists major numerical studies conducted on the FDA
  benchmark and the Reynolds number ($Re_{crit}$) at which the corresponding
  study found flow transition.
  In particular researchers evaluate if their CFD technique can predict the jet
  breakdown location and quantities like pressure and velocity accurately.
  Many studies explore parameters and appropriate mesh densities at which
  their results would match with the experimental data.
  Some of these studies have questioned the suitability of the FDA benchmark
  itself, inquiring if the benchmark should contain more information and be more robust for comparison with CFD. 
  More interestingly some groups have found flow to transition to turbulence at
  $Re_{th}=2000$~\citep{delorme13a, passerini13a, chabannes17a} whereas others
  have found flow transition only at $Re_{th}=3500$~\citep{zmijanovic17a,
  bergersen19a}.

  The study of~\citet{fehn19a} is the only exhaustive work to date in which
  authors applied a high order DG method to study all the $Re_{th}$, and even
  studied additional $Re_{th}$ to explore the $Re_{crit}$ i.e. the critical
  Reynolds number at which the flow would transition in the nozzle.
  Several others demonstrated different jet breakdown locations compared to
  experiments~\citep{passerini13a, bergersen19a}, and some focused on outcomes
  like the \emph{ever changing} location of jet breakdown with increasing
  spatial resolutions~\citep{bergersen19a}.
  Furthermore, it is interesting to note that in some of the previous
  studies~\citep{zmijanovic17a} adding synthetic fluctuations to the inflow was
  necessary to make numerical simulations agree with experiments for the
  $Re_{th}=3500$ case.

  The lattice Boltzmann method (LBM), which is an alternative and relatively
  new technique for the numerical solution of Navier-Stokes equations (NSE) has
  been extensively used in the past decade by fluid dynamics researchers, and
  it has found particular interest in the biomedical community~\citep{sun,
  zhang08a, bernsdorf08a, jain_chiari} as it can represent complex anatomical
  geometries with ease and can enable simulations on massively parallel
  computing architectures~\citep{jain_chiari}.
  Recent works~\citep{jain_chiari, jain19a} applied and validated LBM for
  complex transitional flows in anatomical geometries and found the method
  efficient and valid for such flow regimes.
  Despite pressing needs, however, no extensive effort to the author's
  knowledge has been made to evaluate its efficacy in the aforesaid FDA nozzle
  benchmark.
  On the other hand, due to the diversity of results from different CFD studies
  and varied opinions about the suitability of the FDA benchmark, it becomes
  imperative to explore the suitability of the benchmark itself using a method
  that has not been applied to it before.
  \citet{white11a} did apply LBM to the FDA nozzle benchmark but only for the
  laminar case and their study was more oriented towards exploring the
  suitability of different lattice types.
  The previous works of transitional flow computations using
  LBM~\citep{jain_chiari, jain19a} were also for moderate Reynolds numbers.

  Motivated by the aforementioned needs, this work aims to evaluate if a simple
  LB scheme, without the employment of complex collision models or synthetic
  turbulence at the inflow, will accurately predict the results benchmarked by
  the FDA.
  The focus is on transitional flow regime and thus Reynolds numbers $2000$ and
  $3500$ are only analyzed.
  Physical quantities like \emph{velocity, shear stress} and \emph{pressure}
  and observations like the \emph{jet breakdown} location are compared from
  experiments and simulations.
  Furthermore, insight into questions like \emph{when (Re)}, \emph{where
  (locations)}, \emph{whether} and \emph{how} of flow transition is provided.
  These goals are achieved by conducting LBM based direct numerical simulations
  (DNS) at a very high, and another extreme resolution, which is found to be
  below the scales defined by Kolmogorov theory.
  A comparison against Kolmogorov theory and assessment of Kolmogorov scales is
  further shown to elucidate the role resolutions play in computation of
  complex flow in such a medical device.

\section{Methods}
  \subsection{The FDA nozzle}
  \begin{figure*}%
    \centering%
    \includegraphics[width=\textwidth]{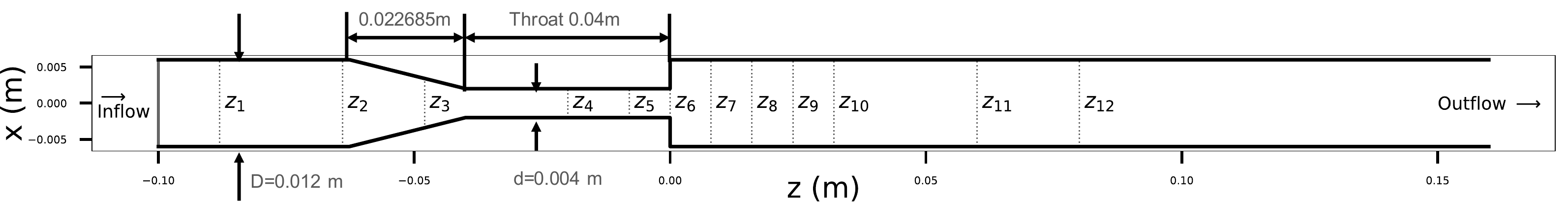}      
    \caption{ Cross sectional view of the FDA nozzle for the sudden expansion case.
    Results were analyzed at various z-locations marked by dotted vertical
    lines to compare against FDA's experimental results: $z_1 = -0.088m,\, z_2
    = -0.064m,\, z_3 = -0.048m,\, z_4 = -0.02m,\, z_5 = -0.008m,\, z_6 =
    0.0m,\, z_7 = 0.008m,\, z_8 = 0.016m,\, z_9 = 0.024m,\, z_{10} = 0.032m,\,
    z_{11} = 0.06m,\, z_{12}
    = 0.08m$. }
    \label{fig:model}
  \end{figure*}

  A cross sectional view of the FDA nozzle is shown in figure~\ref{fig:model}.
  The flow direction is from left to right implying a sudden expansion.
  If the flow is applied from right to left the same geometry will act as a
  canonical diffuser, which is not studied in this work.
  The 3D model of this nozzle was downloaded from the FDA
  website\footnote{\url{https://ncihub.org/wiki/FDA_CFD}}.
  The outlet of the model was extended up to $z=0.253$ m and the inlet was
  extended up to $z=-0.14$ m.
  The radial profiles were recorded at $12$ locations depicted in
  figure~\ref{fig:model} to enable a comparison against the particle image
  velocimetry (PIV) experimental data released by the FDA.

  Simulations were conducted with three different spatial and temporal
  resolutions in order to explore the capabilities of an \emph{off the shelf}
  LBM on a desktop machine to a federal supercomputer.
  Based on previous observations on resolution requirements in LBM simulations
  of transitional flow~\citep{jain16a} a spatial resolution of about $80 \mu m$
  was predicted as a minimum requirement for simulation of $Re_{th}=2000$ and
  accurate analysis of flow characteristics thereof.
  This resolution resulted in $\sim 45$ million lattice sites representing
  fluid inside the nozzle and $50$ lattice cells across the height of the
  nozzle throat.
  This is referred to as normal resolution (NR) hereon.
  For $Re_{th}=3500$ however, the flow was expected to transition and a high
  resolution (HR) of $40 \mu m$ was thus chosen, which resulted in about $357$
  million lattice cells.
  As would be seen in section~\ref{sec:result}, HR resolution sufficed for
  accurate flow simulations for both the Reynolds number.
  To further ensure mesh independence, assess the flow in accordance with
  Kolmogorov theory, and especially because it was not known whether LBM
  simulations will transition to turbulence, an additional set of simulations
  were conducted with an \emph{extreme} resolution (XR) of $20 \mu m$ resulting
  in about $2.88$ billion lattice cells in the fluid domain.
  \begin{table*}
    \centering
    \begin{tabular}{|c c c c c c c|}
    \hline
        & $\delta x (\times10^{-6}m)$ & $\delta t (\times 10^{-6}s)$   & nCells$_{\text{th}}$ & nCells(M)  & nCores     & T(h)  \\ \hline
     NR &   $80$                      & $16$                           & $50$                 & $45$       & $32$       & $192$ \\ \hline     
     HR &   $40$                      & $4$                            & $100$                & $357$      & $38'016$   & $3.5$ \\ \hline     
     XR &   $20$                      & $1$                            & $200$                & $2'880$    & $304'128$  & $12$  \\ \hline
    \end{tabular}
    \caption{Three different spatial and temporal resolutions and the
    corresponding number of lattice cells inside the fluid domain.
    The NR simulations were conducted on $32$ cores of a desktop while others
    on the \emph{SuperMUC-NG}.
    A total of $10$ physical seconds were simulated for all the resolutions. 
    }
    \label{tab:sens}
  \end{table*}
  Table~\ref{tab:sens} lists the employed spatial and temporal resolutions, the
  resulting number of lattice cells, the utilized CPUs and total execution time
  of the simulation.

  \subsection{Setup of a simulation using the lattice Boltzmann method}
    To shed light on the aforementioned choice of resolutions and LB parameters
    a brief explanation is provided here.
    The LBM is based on the mesoscopic representation of movement of fictitious
    particles.
    The particles have discrete velocities and they collide and stream to relax
    towards a thermodynamic equilibrium.  The LB equation recovers the NSE
    under the continuum limits of low Mach and Knudsen numbers.
    Evolution of the particle probability distribution functions over time is
    described by the Lattice Boltzmann equation with the MRT collision matrix:
    \begin{align}
      f_i(\mathbf{r}+\mathbf{c}_i\delta t, t+\delta t) = 
      f_i(\mathbf r,t)
      +  \Omega_{ij} \left( f_i^{e}(\mathbf r,t) - f_i(\mathbf r,t)
      \right)
      \label{eq:LBM}
    \end{align}
    where $f_i$ represents the density distributions of particles which are
    moving with discrete velocity $\mathbf{c}_i$ at a position $\mathbf{r}$ at
    time $\text{t}$.
    The indices which run from $i\!=\!1\!\ldots\!Q$ denote the links per
    element i.e.  the discrete directions, depending on the chosen stencil
    (D3Q19 in our case).
    The collision matrix $\Omega_{ij}$ defines relaxation of various modes of
    the distribution functions $f_i$ towards an equilibrium $f_i^{e}$:
    \begin{align} 
      \label{eq:mainlbe2} 
      f_{i}^{e} = w_i \rho \Bigg(1 + \frac{\mathbf{c}_i \cdot \mathbf{u}}{c_{s}^{2}} -
      \frac{\mathbf{u}^2}{2c_{s}^{2}} + \frac{1}{2} \frac{(\mathbf{c}_i \cdot
      \mathbf{u})^2}{c_{s}^{4}} \Bigg) 
    \end{align}
    where $w_i$ are the weights for each discrete link, $c_s$ is the reference
    speed of sound in LBM obtained by integration of the discrete Boltzmann
    equation along characteristics, and $\mathbf{u}$ is the fluid velocity.  
    The time step in LBM is coupled with the grid size by $\delta t \sim \delta
    x^2$ due to \emph{diffusive} scaling which is employed to recover the
    incompressible NSE.
    Details on the computation of macroscopic quantities from LBM can be found
    elsewhere~\citep{succi}; here we restrict ourselves on the process of
    setting up a flow simulation.

    \subsubsection{Prescription of flow physics}

    Quantities like the mean velocity, $\bar{u}$, on which the Reynolds number
    is based, fluid density $\rho$, and the kinematic viscosity $\nu$ are
    firstly chosen for the fluid to be simulated.
    The Reynolds number is then calculated as:
    \begin{align}
      \label{eq:Re_phy}
      \mathrm{Re}   = \frac{\bar{u} \, d}{\nu}
    \end{align}
    where \textit{d} is the characteristic length equal to the throat diameter
    ($0.004$m) in this case.
    In this study the Reynolds number is based on the nozzle throat.

    The LB method requires the translation of these quantities into lattice
    units.
    Under diffusive scaling, the principal relaxation parameter $\Omega$ is
    fixed and can have a maximum value of $2.0$ for stability.
    The lattice viscosity, $\nu_{lattice}$ is then calculated using the
    relation:
    \begin{align}
      \label{eq:nulattice}
      \nu_{lattice}   = c_s^2 \bigg(\frac{1}{\Omega} - \frac{1}{2} \bigg)
    \end{align}
    where $c_s^2$ is the speed of sound squared at reference state and is equal
    to $\frac{1}{3}$ for the chosen stencil D3Q19.
    The time step is then calculated as:
    \begin{align}
      \label{eq:dt}
      \delta t   = \frac{\nu_{lattice} \, \delta x^2}{\nu}
    \end{align}
    Finally, the lattice velocity is obtained as:
    \begin{align}
      \label{eq:umeanlattice}
      \bar{u}_{lattice}   =  \bar{u} \, \delta t / \delta x 
    \end{align}
    The MRT collision operator allows for different relaxation rates of various
    moments of the distribution function, and it was employed throughout for
    the DNS reported in this manuscript.
    The principal relaxation parameter that defines the kinematic viscosity was
    set to $\Omega=1.90$.
    It is required that the $\bar{u}_{lattice}$ in
    equation~\eqref{eq:umeanlattice} is always less than 0.15 to enforce the
    Mach number limit of the LBM. 
    The $\Omega$ can be adjusted in order to fine tune the lattice velocity and
    beyond its limit, the grid has to be refined (reduce the $\delta x$), which
    is also the principle limitation of the Lattice Boltzmann Method.
    Thus, the $\delta t$ is controlled by the $\delta x$ and the $\Omega$, and
    it is further constrained by the fluid velocity.
    To assert correct prescription of parameters, if the characteristic length
    in equation~\eqref{eq:Re_phy} is replaced by the number of fluid cells
    along that particular characteristic length and $\bar{u}$ and $\nu$ are
    replaced by $\bar{u}_{lattice}$ and $\nu_{lattice}$ respectively, the same
    Reynolds number must be obtained as that obtained from
    equation~\eqref{eq:Re_phy}, which uses physical values. 

    \subsubsection{The simulation framework}

    The employed simulation tool-chain is contained in the end-to-end parallel
    framework APES (adaptable poly engineering simulator)~\citep{apes,
    parco}\footnote{\url{https://apes.osdn.io}}.
    Meshes are created using the mesh generator \emph{Seeder}~\citep{seeder}
    and computations are carried out using the LBM solver
    \emph{Musubi}~\citep{musubi}.
    {\color{black}
      \emph{Musubi} writes out binary files containing physics information to
      the disk.
      These files are then converted to the visualization toolkit (VTK) format
      by the post-processing tool \emph{Harvester}, which is contained within
      the APES framework.
      The open source visualization tool
      Paraview\footnote{https://www.paraview.org} is then used to visualize the
      physics of flow.
      The data for plots is written out by \emph{Musubi} as ASCII files that
      are plotted using the Matplotlib plotting library within the Python
      programming language.
    }

    The 3D model of the nozzle in STL format was read by the mesh generator
    \emph{Seeder} and volume meshes for LBM computations were saved on the
    disk.
    A higher order wall boundary condition described by~\citet{bouzidi13a} was
    prescribed at the walls of the nozzle to reduce the influence of staircase
    artifacts in LBM and ensure rotational symmetry of the setup.
    The \emph{Musubi} LBM solver then computed these meshes (see
    table~\ref{tab:sens}) on the \emph{SuperMUC-NG} petascale system installed
    at the Leibniz Supercomputing Center in Munich, \textsc{Germany}.
    The number of utilized cores on \emph{SuperMUC-NG} ranged from $38\,016$ to
    $304\,128$ (the whole system) based on our previous evaluations which have
    shown that this choice of CPUs results in an optimal utilization of
    compute resources~\citep{parco, parco16}.

    \subsubsection{Initial and boundary conditions}
    Initial conditions were set to zero pressure and velocity and a no-slip
    boundary condition was maintained at the walls.
    The flow rates at inlet were characterized based on the throat Reynolds
    number.
    The fluid itself was chosen to be Newtonian, with a fluid density of $1056
    kg/m^3$ and dynamic viscosity of $0.0035 N s/m^2$.

    A parabolic velocity profile was prescribed at the inlet by quadratically
    interpolating the velocity at lattice nodes, and a zero pressure was
    maintained at the outlet.
    No synthetic turbulence was prescribed at the inlet unlike some previous
    studies.
    The outflow in such flow regimes can be critical resulting in back flow or
    instabilities in the LBM algorithm for which an extrapolation boundary
    condition was employed~\citep{junk2011}.
    
    Following \citet{fehn19a} the simulated time interval was chosen
    on the basis of a characteristic flow-through time.
    The length of the throat, $L_{th}$ section and the mean velocity, $\bar{u}$
    were chosen as characteristic length scale and the reference velocity
    respectively.  This resulted in a flow through time given by:
    \begin{align} \label{eq:ft}
        T_{ft} = \frac{L_{th}}{\bar{u}}   
    \end{align}
    For the $Re_{th}=2000$ case this time was approximately $0.21$ s whereas
    for the $Re_{th}=3500$ case it was $0.12$ seconds.
    Flow was allowed to develop for initial $2$ seconds implying about $10$ and
    $20$ flow throughs for the $Re_{th}=2000$ and $Re_{th}=3500$ case
    respectively.
    Within the computational bounds a total of $10$ physical seconds were
    simulated for each case.
    A total of $50000$ samples were gathered every second for the analysis of
    flow characteristics.

    \subsection{Flow characterization}

    For the analysis of turbulent characteristics, the three dimensional
    velocity field was decomposed into a mean and fluctuating component as:
        \begin{align} \label{eq:trip_decomp}
          u_i(x,t) = \bar{u}_i(x) + u_{i}^{\prime}(x,t)
        \end{align}
    The Turbulent Kinetic Energy (TKE) was derived from the fluctuating
    components of the velocity in three directions as:
    \begin{align} \label{eq:tke}
      k = \frac{1}{2}\Big( {u_{x}^{\prime 2} + u_{y}^{\prime 2} + u_{z}^{\prime 2}} \Big)
    \end{align}
    The dimensionless Strouhal number, St is defined as:
    \begin{align} \label{eq:st}
        St = \frac{f d}{\bar{u}}
    \end{align}
    where f is the frequency of flow fluctuations, d and $\bar{u}$ are the
    characteristic length and mean velocities.
    A power spectral density of the TKE was computed as a function of the
    Strouhal number at various centerline locations using the Welch's
    periodogram method to analyze the inertial and viscous subranges.

    \subsection{Kolmogorov microscales} \label{subsec:klmgrv}
    The smallest structures that can exist in a turbulent flow can be estimated
    on the basis of Kolmogorov's theory~\citep{pope} {\textendash} used in this
    work to assess the quality of the resolutions in DNS.
    Viscosity dominates and the TKE is dissipated into heat at the Kolmogorov
    scale~\citep{pope}.
    The Kolmogorov theory, in general, applies to fully developed turbulent
    flows with Reynolds numbers much higher than those studied in this work.
    The reference to Kolmogorov scales and theory here is thus indicative for
    the assessment of mesh independence as has been done in various such
    studies at low Reynolds numbers~\citep{jain19a, helgeland14a}.

    The Kolmogorov scales, non-dimensionalized with respect to the velocity
    scale $\bar{u}$ and the length scale d (throat diameter) are computed from
    the \emph{fluctuating} component of the non-dimensional strain rate defined
    as:
          \begin{align} \label{eq:flstr}
            s^{\prime}_{ij} = \frac{1}{2}\bigg(\frac{\partial u_{i}^{\prime}}{\partial x_j} +
            \frac{\partial u_{j}^{\prime}}{\partial x_i}\bigg) 
          \end{align}
    The Kolmogorov length, time and velocity scales are then respectively
    computed as:
        \begin{align} \label{eq:eta}
          \eta  = \bigg( \frac{1}{Re^2} \frac{1}{2 s^{\prime}_{ij}s^{\prime}_{ij}} \bigg)^{1/4}
        \end{align}
        \begin{align} \label{eq:taueta}
          \tau_{\eta} = \bigg( \frac{1}{2 s^{\prime}_{ij}s^{\prime}_{ij}}\bigg)^{1/2}     
        \end{align}
        \begin{align} \label{eq:ueta}
          u_{\eta} = \bigg( \frac{2s^{\prime}_{ij}s^{\prime}_{ij}}{Re^2} \bigg)^{1/4}
        \end{align}
    Based on these scales, the quality of the spatial and temporal resolutions
    employed in a simulation can be estimated as the ratio of $\delta x$ and
    $\delta t$ against the corresponding Kolmogorov scales i.e.
          \begin{align} \label{eq:lp}
            l^{+} = \frac{\delta x}{\eta}
          \; \hspace{10pt}
            t^{+} = \frac{\delta t}{\tau_{\eta}}
          \end{align}
    The Kolmogorov scales in this study were computed along various locations at the centerline.
    The fluctuating strain rate was averaged between locations $z_8$ and
    $z_{12}$ (figure~\ref{fig:model}) due to variations in TKE caused by jet
    breakdown in these locations, and the Kolmogorov scales were computed from
    this averaged strain rate.

    \subsection{Comparison to experiments}
      
      The radial velocity profiles in the streamwise direction and the mean
      centerline velocity at $12$ locations shown in figure~\ref{fig:model}
      were plotted against corresponding data from $5$ PIV experiments.
      Instantaneous centerline velocities and pressure were analyzed at $120$
      locations each $0.002$ m apart along the length of the nozzle.
      The $5$ FDA experiments were conducted by $3$ independent laboratories
      where one laboratory ran three trials resulting in $5$ data
      sets~\citep{hariharan11a}.
      The axial component of the velocity $u_z$ was plotted directly against
      the data from experiments\footnote{Note that other works~\citep{fehn19a,
      passerini13a} have normalized the $u_z$ with respect to $\bar{u}_{in}$}.
      
      Pressure was probed and averaged at $12$ circular cross sections across
      the length of the nozzle ($z_1\,-\,z_{12}$).
      To enable comparison against experiments, the mean pressure difference
      was normalized with respect to the average velocity at the throat
      $\bar{u}$:
      \begin{align}
            \Delta p^{norm} = \frac{p_z - p_{z_0}}{0.5 \rho \bar{u}^{2}}
      \end{align}
      where $\rho$ is the fluid density and $\bar{u}$ is the mean velocity
      at the nozzle throat, on which the $Re_{th}$ was based
      (see equation~\eqref{eq:Re_phy}).

      The shear stress was computed from the LBM simulations as it is one of
      the quantities of interest to predict hemolysis, and was directly
      compared to the experimental data.

      {\color{black}
        The differences between a particular LBM simulation case and an
        experimental dataset were quantified on the basis of simple relative
        percentage errors, computed as:
        \begin{align} \label{eq:err}
          \delta = \left|~\frac{U_{ref} - U_h}{U_{ref}}~\right| \times 100
        \end{align}
        where $U_{ref}$ denotes the reference solution (experiments in our
        case) and $U_h$ denotes the LBM computed solution; the corresponding
        solutions can be for pressure, velocity or shear stress.
      }

\section{Results}       \label{sec:result}
  The flow transitioned to turbulence at $Re_{th}=2000$ and it became fully
  turbulent at $Re_{th}=3500$.
  The NR simulation was stable for $Re_{th}=2000$ during the whole $10$ seconds
  that were simulated.
  For $Re_{th}=3500$, however, the NR simulation crashed as soon as the flow
  reached the outflow due to the instabilities in that region, and local
  fluctuations in the Mach number limit of the LB algorithm
  (equation~\ref{eq:umeanlattice}).

  \subsubsection*{Kolmogorov microscales}
  To assess the mesh independence of the simulations we start with an analysis
  of the Kolmogorov microscales, shown for different resolutions and $Re_{th}$
  in table~\ref{tab:klmgrv}.
  \begin{table*}
    \begin{tabular}{|c|p{1.2cm}p{1.2cm} p{1.2cm} p{1.2cm} p{1.2cm}  |p{1.2cm} p{1.2cm} p{1.2cm} p{1.2cm} p{1.2cm}|}
    \hline
           &\multicolumn{5}{c|}{$Re_{th}=2000$}& \multicolumn{5}{c|}{$Re_{th}=3500$}                   \\  \cline{2-11}
           & $\eta (\mu m)$   & $\tau_{\eta} (\mu s)$   & $u_{\eta} (m/s)$  & $l^{+}$   & $t^{+}$ 
           & $\eta (\mu m)$   & $\tau_{\eta} (\mu s)$   & $u_{\eta} (m/s)$  & $l^{+}$   & $t^{+}$       \\  \hline
    NR     & $4.94$           & $8.37$                  & $1.19$            & $16.19$   & $1.91$ 
           & $\Theta$         & $\Theta$                & $\Theta$          & $\Theta$  & $\Theta$      \\  \hline
    HR     & $16.32$          & $8.69$                  & $1.97$            & $2.45$    & $0.46$ 
           & $12.57$          & $4.08$                  & $3.80$            & $3.18$    & $0.98$        \\  \hline 
    XR     & $21.97$          & $7.14$                  & $2.10$            & $0.91$    & $0.14$ 
           & $18.86$          & $4.34$                  & $4.10$            & $1.06$    & $0.23$        \\  \hline 

    \end{tabular}
    \caption{Kolmogorov scales computed from NR, HR and XR resolutions for
    $Re_{th}=2000$ and $Re_{th}=3500$.}
    \label{tab:klmgrv}
  \end{table*}
  The $l^{+}$ from NR resolution for $Re_{th}=2000$ is $16.19$ thus clearly
  indicating this resolution as insufficient for accurate assessment of
  turbulent characteristics.
  These scales from the HR simulations for both $Re_{th}$ are sufficiently
  close to the corresponding Kolmogorov scales while from XR simulations they
  attain a value equal to the Kolmogorov scales.
  As observed later, the XR resolutions, being resolved exactly at the
  Kolmogorov length scales result in minor deviances in flow characteristics if
  compared to HR resolutions.
  The $\tau_{\eta}$ are $<1$ in most of the cases due to the small time step of
  LB simulations.

  \subsubsection*{Jet breakdown location}
  \begin{figure*}%
    \centering%
    \subfloat[$Re_{th}=2000$ NR]{
      \includegraphics[width=\textwidth]{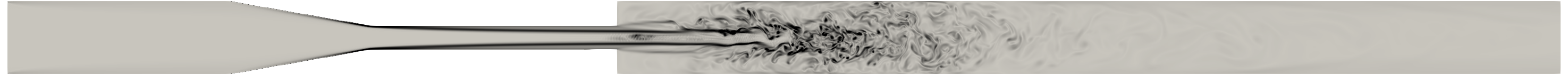}   \label{fig:2knr}
    }\\
    \subfloat[$Re_{th}=2000$ HR]{
      \includegraphics[width=\textwidth]{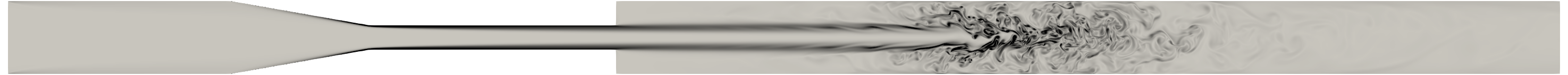}   \label{fig:2khr}
    }\\
    \subfloat[$Re_{th}=2000$ XR]{
      \includegraphics[width=\textwidth]{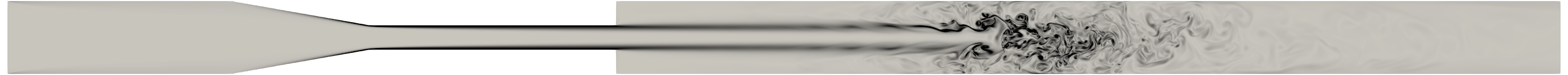}   \label{fig:2kxr}
    }\\
    \subfloat[$Re_{th}=3500$ HR]{
      \includegraphics[width=\textwidth]{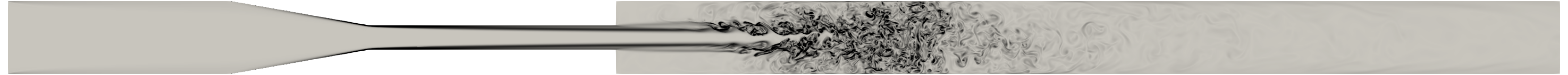}   \label{fig:35hr}
    }\\
    \subfloat[$Re_{th}=3500$ XR]{
      \includegraphics[width=\textwidth]{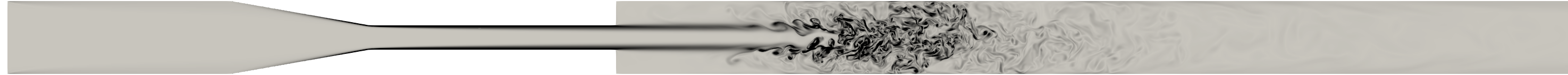}   \label{fig:35xr}
    }
    \caption{Snapshots of the instantaneous vorticity during t=10 second of
    the simulation across a bisecting plane in the FDA nozzle from HR and XR
    LBM simulations for $Re_{th}=2000$ and $Re_{th}=3500$.
    The vorticity magnitude is scaled according to $Re_{th}$ and ranges from
    $0-2Re_{th}$.
    }
    \label{fig:vort}
  \end{figure*}

  \begin{figure*}%
    \centering%
    \subfloat[$Re_{th}=2000$]{
      \includegraphics[width=\textwidth]{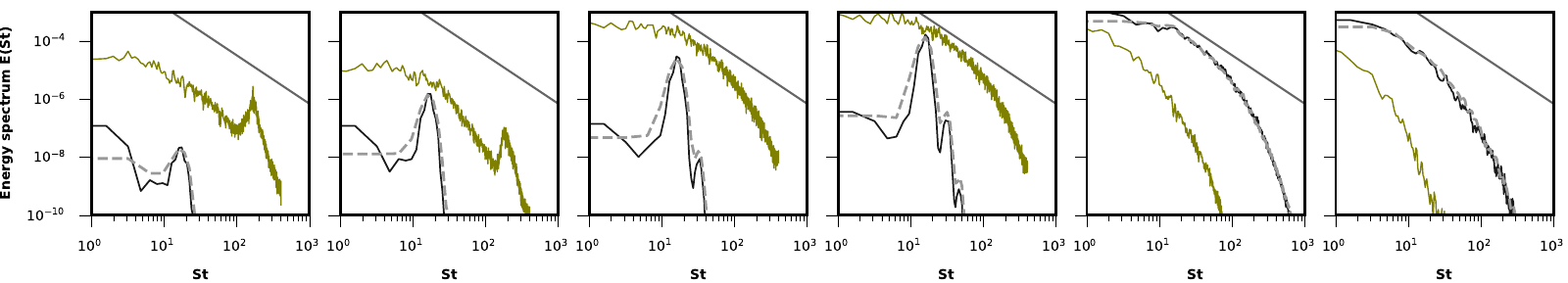} \label{fig:2kpsd}
    }\\
    \subfloat[$Re_{th}=3500$]{
      \includegraphics[width=\textwidth]{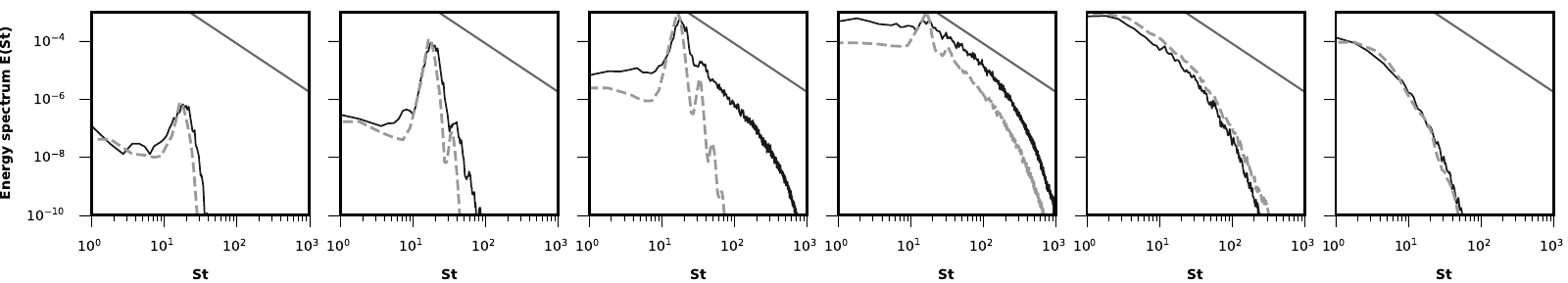} \label{fig:35psd}
    }
    \caption{Spectral density of the turbulent kinetic energy at $6$ locations
    along the centerline after the sudden expansion ($z_7-z_{12}$ in
    figure~\ref{fig:model}) along the centerline is plotted against the
    Strouhal number (St).
    The PSD is computed during the last $8$ seconds of the simulations using $4
    \times 10^5$ time steps in total.
    Dark and dotted gray lines show the PSD respectively from HR and XR LBM
    simulations.
    For $Re_{th}=2000$ PSD from NR simulations is shown in green line.
    The solid gray line shows the Kolmogorov's $\frac{-5}{3}$ decay.
    }
    \label{fig:psd}
  \end{figure*}

  The regions of flow breakdown and maximum turbulent activity are highlighted
  in figure~\ref{fig:vort}, which shows the instantaneous vorticity magnitude
  at the $t=10$ second of simulations.
  Jet breakdown location identified from NR simulations of $Re_{th}=2000$ lies
  between $z_9$ and $z_{10}$.
  A secondary flow jet however starts emanating right after the expansion.
  It may be concluded that the NR resolution captures the onset of turbulence
  but with a largely compromised accuracy as the location of jet breakdown is
  different from all the previously reported experiments~\citep{stewart12a}.
  On the other hand the vorticity plots for $Re_{th}=2000$ from both HR and XR
  LBM simulations demonstrate that the location of jet breakdown is more or
  less identical (between $z_{11}$ and $z_{12}$) and resolutions do not seem to
  influence the location of jet breakdown.
  The jet breakdown location is between $z_9$ and $z_{10}$ at $Re_{th}=3500$
  from HR simulations while it is shifted further downstream by approximately
  half of the throat diameter from XR simulations.
  It is emphasized that these are \emph{instantaneous} vorticity plots during
  the t=10 second of the simulation as due to immense memory requirements an
  ensemble averaging was not possible.
  Furthermore, it is expected that an ensemble average of at least $10\,000$
  snapshots would be needed to have an accurate assessment of jet breakdown
  location.

  In the supplementary material, animations of vorticity and velocity field for
  both $Re_{th}$ from HR simulations are provided over the last one second
  ($9-10$) of the simulations.
  It may be observed from the animations that at $Re_{th}=2000$ the jet itself
  experiences distortion over time course of the simulation.
  If we take a closer look at the figure~\ref{fig:2khr} we can see development
  of \emph{discontinuities} over the jet after $3$ nozzle diameters downstream
  of expansion.
  When the flow breaks down, vortices merge, annihilate and interact with the
  jet to \emph{distort} the jet itself over the course of time and resolution
  does not seem to play a role here.
  This circumstance is in fact natural for a transitional flow as it is not
  fully developed turbulence but the jet is between laminar and transitional
  regime.
  For $Re_{th}=3500$, on the other hand, the jet does not get distorted over
  the course of time but it shifts further downstream at XR resolution
  (figure~\ref{fig:35xr}).
  In this case, a shear layer develops around the jet itself as can be clearly
  seen from the animations as well as the instantaneous snapshots of vorticity.
  A further grid refinement for the case of $Re_{th}=3500$
  (figure~\ref{fig:35xr}) resolves this shear layer better to shift the jet
  breakdown location further downstream by about half a throat diameter
  ($d/2=0.002m$) as this is a fully developed turbulent flow field.

  An accurate assessment of turbulent activity at various locations, and from
  different resolutions can be obtained from the PSD plots of
  figure~\ref{fig:psd}, computed at locations downstream of the expansion
  ($z_7-z_{12}$ from figure~\ref{fig:model}).
  The PSD is computed over $8$ seconds of the simulation to obtain abundant
  statistics and overcome previous issues of insufficient averaging.
  The dark and dotted gray lines respectively show PSD from HR and XR LBM
  simulations whereas the solid gray line indicates Kolmogorov's $\frac{-5}{3}$
  decay.
  For $Re_{th}=2000$ the green line shows the PSD from NR simulations, which
  corroborates previous observations from vorticity plots and Kolmogorov scales
  and confirms the inadequacy of this resolution.
  For the $Re_{th}=2000$ case it is seen that the flow transitions at locations
  $z_{11}$ and $z_{12}$ and the turbulent activity captured by HR and XR
  simulations is the same with a few frequencies in the inertial subrange.
  The third plot of figure~\ref{fig:35psd} is the most enlightening about the
  jet breakdown location for $Re_{th}=3500$ from HR and XR simulations and
  corroborates the observations from plot~\ref{fig:35cent}.
  Clearly the jet does not break down at location $z_9$ in XR simulations as
  the spectrum falls off rapidly whereas in HR simulations the spectrum tends
  to attain a Kolmogorov like decay~\citep{jain19a} at this location.
  The following figures then substantiate this observation whereby at $z_{10}$
  the jet is already turbulent from both HR and XR simulations and turbulent
  activity becomes similar beyond this (last two plots of
  figure~\ref{fig:35psd}).

  \subsubsection*{Comparison of velocity, pressure and shear stress against experiments}

  \begin{figure*}%
    \centering%
    \subfloat[$Re_{th} = 2000$]{
      \includegraphics[width=0.50\textwidth]{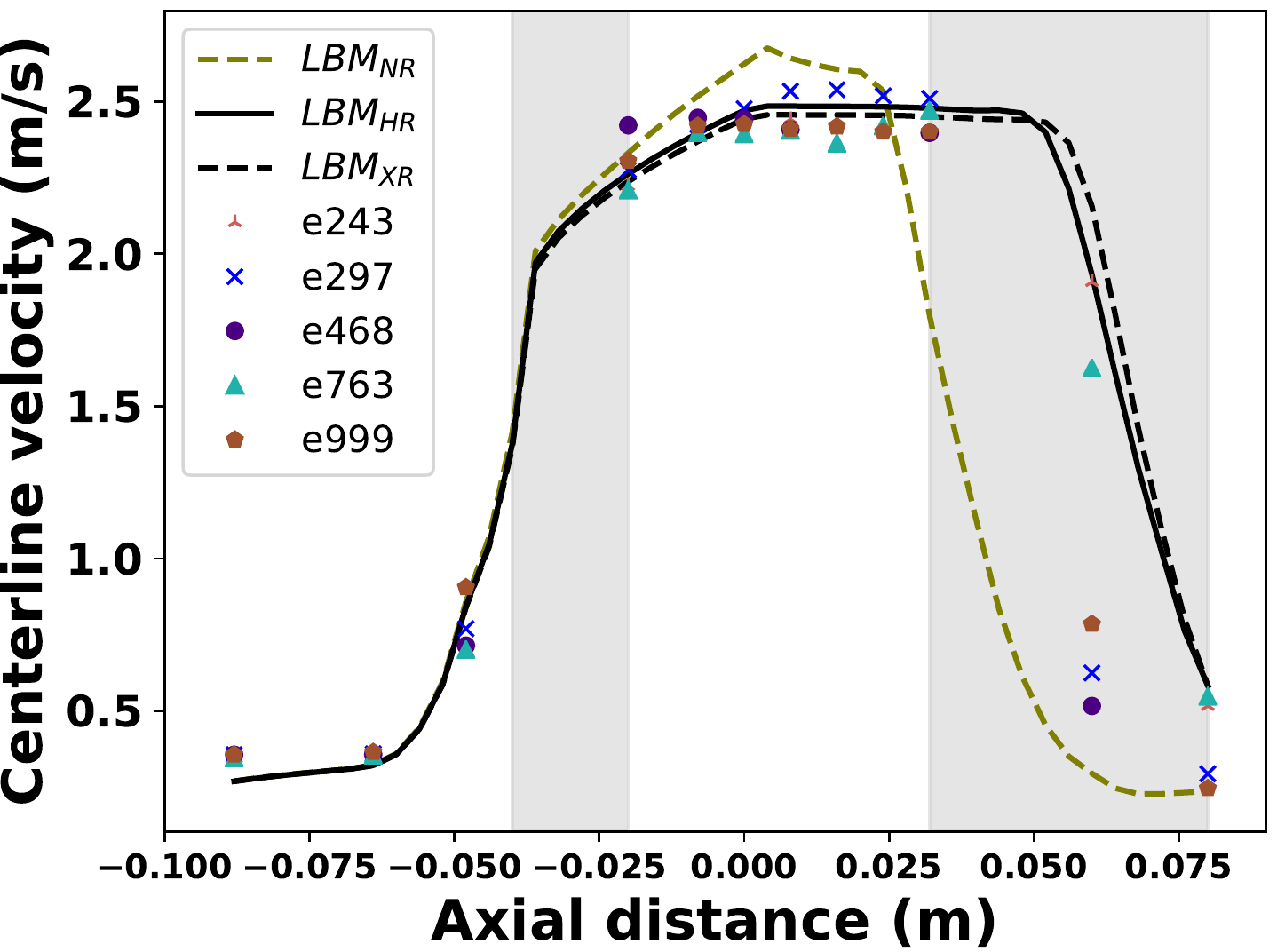} \label{fig:2kcent}
    }
    \subfloat[$Re_{th} = 3500$]{
      \includegraphics[width=0.50\textwidth]{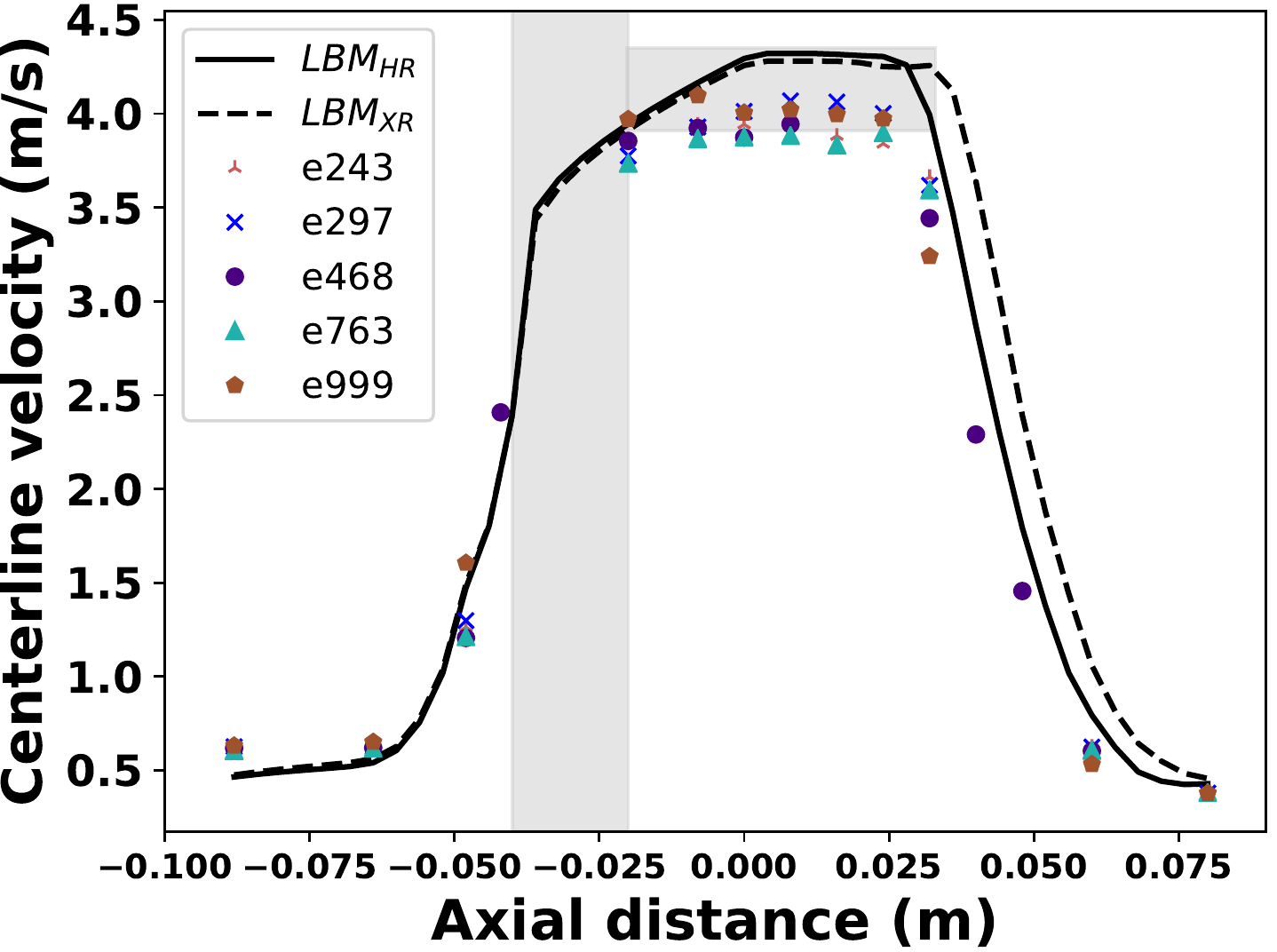} \label{fig:35cent}
    }
    \caption{Time averaged centerline velocity at $86$ locations along the
    length of the nozzle for two different Reynolds numbers and all the
    employed LBM resolutions.
    The centerline velocity is compared with $12$ ($z_1$ -- $z_{12}$ from
    figure~\ref{fig:model}) experimental data points along the centerline
    available from $5$ distinct PIV experiments.
    The number in the legend corresponds to the experiment ID from the FDA
    website.
    The gray bars highlight the regions with largest discrepancy between
    experiments and simulations.
    At $Re_{th}=3500$ the NR simulation was unsuccessful.
    }
    \label{fig:center}
  \end{figure*}
  \begin{figure*}%
    \centering%
    \subfloat[$Re_{th} = 2000$]{
      \includegraphics[width=0.5\textwidth]{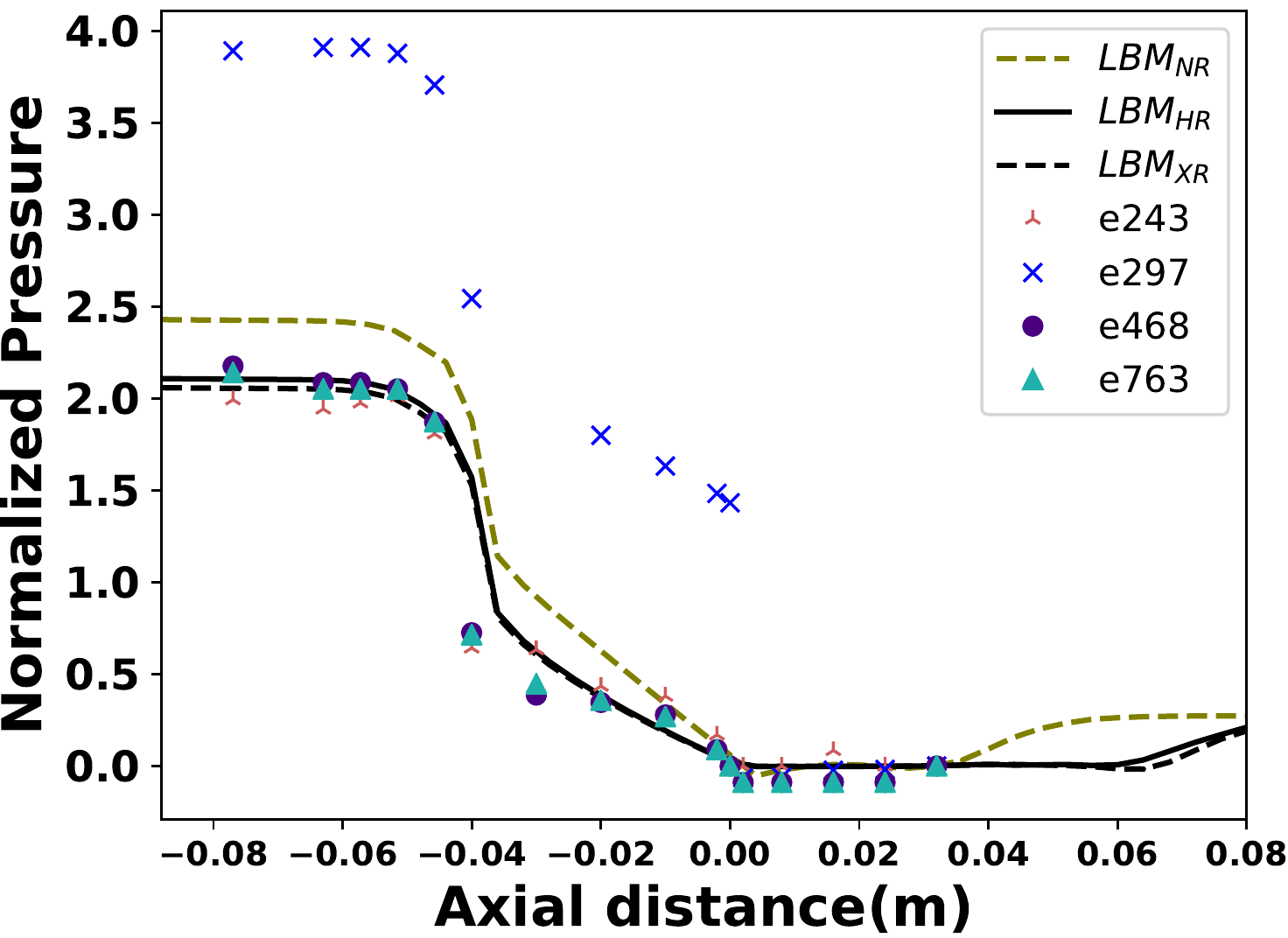} \label{fig:2kpres}
    }
    \subfloat[$Re_{th} = 3500$]{
      \includegraphics[width=0.5\textwidth]{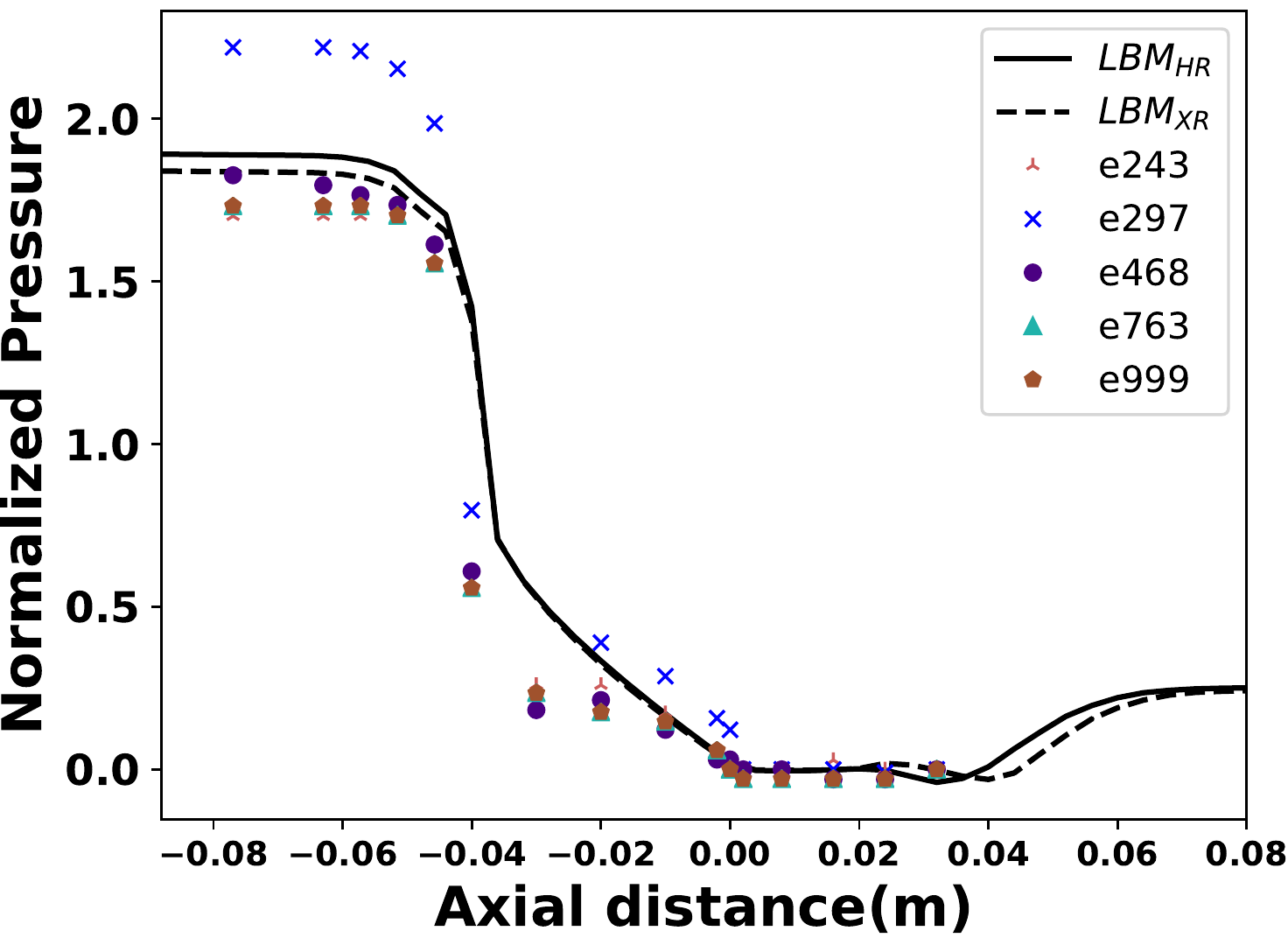} \label{fig:35pres}
    }
    \caption{Time averaged centerline pressure normalized by the mean throat
    velocity $u_{th}$ versus the axial distance computed from two sets of LBM
    resolutions is compared against pressure data from $5$ PIV experiments of
    the FDA for two different Reynolds numbers.
    Experimental data is plotted at $12$ locations along the centerline ($z_1$
    -- $z_{12}$ in figure~\ref{fig:model}) for experiments whereas $86$ points
    between these locations are plotted from simulations.
    Note that for $Re_{th}=2000$ case the pressure data from experiment $\#999$
    is not available. 
    At $Re_{th}=3500$ the NR simulation was unsuccessful.
    }
    \label{fig:press}
  \end{figure*}

  Figure~\ref{fig:center} shows the averaged centerline velocity computed at
  $86$ axial locations between $z_1$ and $z_{12}$ for both the Reynolds numbers
  from HR and XR LBM simulations and additionally from NR simulations for the
  $Re_{th}=2000$ case.
  The corresponding data from $5$ experiments at $12$ locations is plotted for
  comparison.
  The shaded bars in the plots highlight the regions of largest discrepancy
  between experiments and simulations.
  As can be seen, for $Re_{th}=2000$ there is a significant difference between
  experiments and simulations between locations $z=-0.04$ and $z=-0.02$, which
  is the region just after the end of gradual contraction, or the beginning of
  the nozzle throat.
  The velocity matches well at the latter part of the throat and this difference
  is seen again in the jet breakdown location, which matches reasonably with one
  experiment only ($e243$) for both the HR and XR resolutions.
  {\color{black}
    The difference in velocity in regions between $z=0.06$ and $z=0.07$ is
    about $10\%$ between few experiments and simulations, computed using
    equation~\eqref{eq:err}.
  }
  From NR simulations the jet breakdown location is completely different from
  experiments and HR and XR simulations as was seen in previous plots.

  For the $Re_{th}=3500$ case similar trends in the throat area are seen while
  the jet breakdown location from HR simulations matches more closely to the
  experiments.
  This location from the XR simulations is comparatively further away by half
  the throat diameter.
  {\color{black}
    The velocities estimated by the LBM at the throat are also higher by about
    $6-9\%$ than those from the experiments.
    For $Re_{th}=3500$, the velocities computed by LBM right after the
    expansion at locations around $z=0.025$ are overestimated by about $10\%$
    compared to the experiments.
  }

  The cause of discrepancies in the starting of the throat area (shaded
  regions) cannot be unequivocally stated as the data points available from the
  experiments are very few.
  In the plot from the simulations, there are $14$ samples between locations
  $z_3$ and $z_4$ (figure~\ref{fig:model}).
  If the data points between these two locations were omitted, the curves would
  look like a straight line as in the experiments and previously reported
  studies~\citep{fehn19a}.
  The pressure at the throat right after the contraction reduces considerably
  and the minor discontinuities at the corners might explain these differences.

  This is further elaborated in the centerline average pressures plotted in
  figure~\ref{fig:press}.
  Qualitatively, similar trends from experiments and simulations are observed,
  and the difference between HR and XR simulations for both the Reynolds number
  is much smaller.
  {\color{black}
    Interestingly, the pressure drop for $Re_{th}=3500$ at locations before the
    expansion ($z<0.0$) is overestimated by about $5-6\%$ and matches only one
    experiment.
  }
  The data points for pressure from PIV experiments are relatively few
  disabling a better and insightful comparison.
  Furthermore because the outflow length in LBM simulations is larger than that
  in experiments, it may account for minor differences in pressure changes.

  \begin{figure*}%
    \centering%
    \subfloat[$Re_{th} = 2000$]{
      \includegraphics[width=\textwidth]{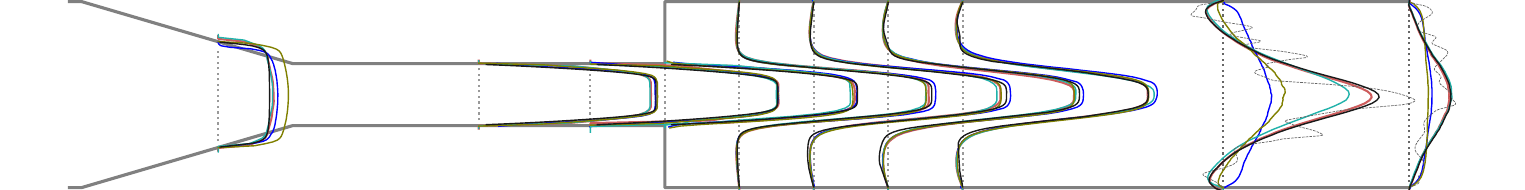} \label{fig:2krad}
    }\\
    \subfloat[$Re_{th} = 3500$]{
      \includegraphics[width=\textwidth]{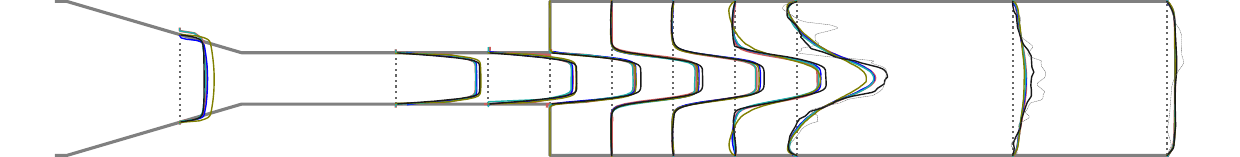} \label{fig:35rad}
    }
    \caption{Radial velocity profiles at $10$ locations ($z_3-z_{12}$ shown in
    figure~\ref{fig:model}) along the length of the nozzle for two different
    Reynolds numbers averaged over $8$ seconds or $400 000$ time steps.
    The velocities are normalized to enable an intuitive comparison.
    The black lines show the averaged velocities computed from XR LBM
    simulations whereas red, blue, green and olive colored lines are the data
    from experiment ids $\#243$, $\#297$, $\#763$ and $\#999$ respectively.
    The gray dotted lines at locations $z_{11}$ and $z_{12}$ for $Re_{th}=2000$
    and at $z_{10}$, $z_{11}$ and $z_{12}$ for $Re_{th}=3500$ show
    instantaneous radial velocities from LBM simulations to depict the location
    of jet breakdown.}
    \label{fig:radial}
  \end{figure*}

  \begin{figure*}[h]%
    \centering%
    \subfloat[$Re_{th} = 2000$]{
      \includegraphics[width=\textwidth]{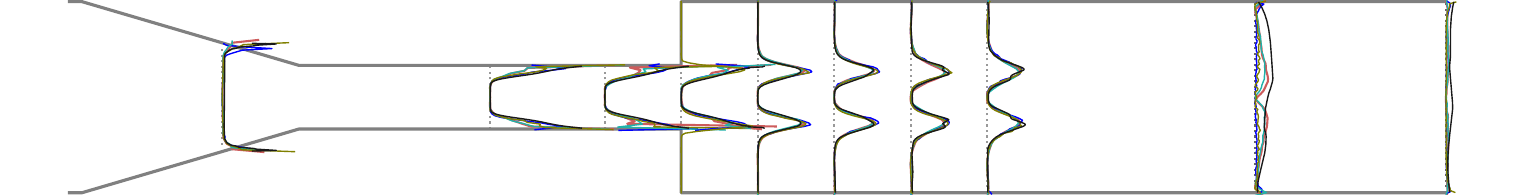} \label{fig:2kwss}
    }\\
    \subfloat[$Re_{th} = 3500$]{
      \includegraphics[width=\textwidth]{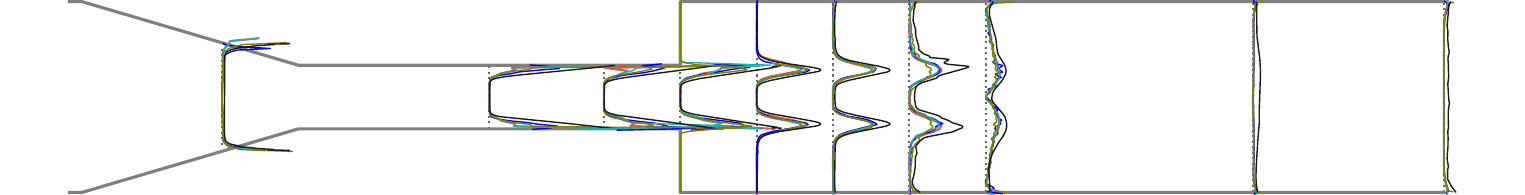} \label{fig:35wss}
    }
    \caption{Radial shear stress profiles at $10$ locations ($z_3-z_{12}$ shown
    in figure~\ref{fig:model}) along the length of the nozzle for two different
    Reynolds numbers averaged over $8$ seconds or $400 000$ time steps.
    The shear stresses are normalized to enable an intuitive comparison.
    The black lines show the averaged velocities computed from XR LBM
    simulations whereas red, blue, green and olive colored lines are the data
    from experiment ids $\#243$, $\#297$, $\#763$ and $\#999$ respectively.  }
    \label{fig:wss}
  \end{figure*}

  Figure~\ref{fig:radial} shows the radial profiles of mean streamwise velocity
  from XR LBM simulations at $10$ locations after the contraction ($z_3-z_{12}$
  from figure~\ref{fig:model}).
  The ensemble averages are gathered over the last $8$ seconds of the
  simulation or for $4 \times 10^5$ time steps.
  {\color{black}
    In this case the velocities have been normalized by $\bar{u}$ for a direct
    intuition about the differences in experiments and simulations.
    The red, blue, green and olive colored lines are the data from experiment ids
    $\#243$, $\#297$, $\#763$ and $\#999$ respectively\footnote{Plots for
    experiment $\#468$ have been omitted as several radial profiles are not
    available from the data of this experiment.}.
  }
  At locations $z_{11}$ and $z_{12}$ for $Re_{th}=2000$ and at $z_{10}-z_{12}$
  for $Re_{th}=3500$, the instantaneous radial profiles at t=10 second of the
    simulation are plotted in gray dotted lines to depict regions where the
    flow jet broke down.
  For the $Re_{th}=3500$ case the velocity from LBM in the post expansion
  region is higher than the experiments mainly in the jet core region as was
  also observed in figure~\ref{fig:35cent}.
  Except for the jet breakdown locations, a reasonable agreement in radial
  profiles can be seen against all the experiments. 
  As has been observed throughout this study the interlaboratory variations in
  the experiments are quite large~\citep{stewart12a} making a quantitative
  comparison more difficult.

  Shear stress from experiments and computations is compared in
  figure~\ref{fig:wss}.
  Here the shear stress has been normalized by mean shear stress so that a
  direct comparison can be enabled.
  Excellent agreement for both the Reynolds number can be observed in mean
  radial velocities as well as the shear stress\footnote{The shear stress at
  $z_3$ seems to \emph{go out} of the nozzle. This is a visual artifact as the
  shear rises dramatically at the walls.}.
  The shape of the shear stress profile at locations before the expansion
  ($z<0.0$) is flatter in the simulations whereas in experiments it exhibits a
  skewed shape.
  Other minor differences lie in the regions of jet breakdown and are a
  consequence of the fact that $4 \times 10^5$ sample points are used for
  gathering statistics, while good, are not perfect for the reproduction of a
  converged state.
  Despite the computational resources deployed for this study, sampling more
  statistics was considered prohibitive.
  For $Re_{th}=3500$, LB overestimates the velocity and the shear stress at the
  location of jet breakdown, which was also observed in
  figure~\ref{fig:center}.
  {\color{black}
    These differences are about $3-4\%$ are mostly confined around the jet
    breakdown locations.
  }
  %

\section{Discussion}
  It is the first time that LBM has been applied to study transitional flows in the two decade old FDA benchmark.
  The study has found that LBM can predict the jet breakdown location and compute macroscopic quantities to a reasonable accuracy for a transitional flow in a biomedically relevant device.
  In addition to a comparison the main physical insight is the distortion in jet for transitional flow case of $Re_{th}=2000$ and assessment of the Kolmogorov microscales.
  Here we discuss the agreement and discrepancy in results as well as implications of this study.

  \subsubsection*{Analysis of the flow characteristics and comparison to previous works and experiments}
  Qualitatively the LB computed flow transitioned to turbulence at
  $Re_{th}=2000$ as it did in most of the experiments~\citep{stewart12a}, and flow
  field at $Re_{th}=3500$ was reminiscent of a partially developed turbulence
  indicated by the immense chaotic activity.
  Quantitatively, a satisfactory match between experiments and computations was observed in averaged velocities, shear stress, and pressure as well as the jet breakdown locations.
  This agreement has been obtained without parameter tuning in the method, and in particular without adding synthetic fluctuations as has been necessary for other studies~\citep{zmijanovic17a}.
  The centerline velocity and pressures computed from LBM had marked differences, in particular in the throat and the expansion area.
  In particular for the $Re_{th}=3500$ case the LB computed velocities were higher by about $10\%$ in comparison to the experiments.
  While an exact reason for this is not known it may be observed that the LB computed radial velocity profiles (figure~\ref{fig:35rad})
  had a more pronounced recirculation region and a higher velocity only in the jet core {\textendash} not seen in the experiments.
  
  A mesh sensitivity analysis revealed that resolutions within an order of magnitude of the Kolmogorov microscales are necessary to accurately capture the flow field
  and a further refinement down to the Kolmogorov scales results in slightly distal jet breakdown of flow downstream of expansion for $Re_{th}=3500$ while the averaged macroscopic quantities are not much affected.
  The NR resolution failed for the higher $Re_{th}=3500$ and the results were erroneous for $Re_{th}=2000$. 
  Previous studies have found a much larger dependence of results on the mesh density.
  \citet{delorme13a} used LES model and found good match except that their simulations found relatively less turbulent nature of the flow attributed to grid resolutions.
  \citet{passerini13a} performed a number of simulations for each $Re$ they studied (table~\ref{tab:fdastud}) to arrive at an optimal grid resolution.
  \citet{zmijanovic17a} used three mesh resolutions in their LES simulations and found excellent match with experiments from the highest resolution of $50M$ cells.
  In the present LBM simulations, going from HR to XR increased the computational effort remarkably and brought the resolutions right down to the Kolmogorov microscales.
  The benefit of this was pronounced at $Re_{th}=3500$ while at $Re_{th}=2000$ almost no advantage in the results was seen from HR to XR.
  The HR setup is thus adequate in this configuration as suggested by previous studies~\citep{jain16a} as well.
  The scalability of LBM to XR scale may be leveraged in future to incorporate
  physics of red blood cells or other particles in the blood~\citep{sun} to
  answer relevant questions in physiology.

  In the present work the role of boundary conditions has not been explored in detail and the prescription of inflow and outflow boundary conditions is likely to influence the results.
  \citet{passerini13a} extended the in- and outflow according to the $Re_{th}$
  and the recent precursor approach adopted by~\citet{fehn19a} seems to
  overcome most of the effects of inflow pipe length.
  When a simulation is refined, the accuracy of the boundary conditions also increases, which explains the \emph{eternal} change in solution upon increasing resolutions.
  In particular the disturbances that emanate from the outflow are reduced upon refining the mesh and time step, which would explain the relatively further breakdown of jet downstream of the expansion.

  \subsubsection*{Jet breakdown location}
  A number of previous studies have found the jet breakdown location sensitive to parameters used in CFD~\citep{passerini13a, zmijanovic17a}.
  Here, as noted above there were no parameters that were varied in the LBM
  simulations except for the grid refinement and the corresponding time step
  size.
  For $Re_{th}=2000$, we noted that the jet breakdown location was largely the same from HR and XR resolutions.
  A new observation, seemingly overlooked in previous studies was the propagating distortion in the jet at $Re_{th}=2000$ (animations in supplementary information).
  As the flow was just transitioning, the jet lost its momentum at times, tended to restabilize the flow, and then gained momentum again resulting in shifts in the breakdown location.
  For $Re_{th}=3500$ the jet broke down half a nozzle diameter further downstream of the expansion in XR simulations compared to HR.
  The reason for that is the better accuracy at the sudden expansion, which stabilized the perturbances in that area, and their propagation thereof.
  Nonetheless, the breakdown location from XR simulations was just half the throat diameter, $d/2$, downstream, and its relevance can be queried.
  More interestingly the flow restabilization regions from both the resolutions were the same for $Re_{th}=3500$, and the XR resolution thus narrowed down the length of chaotic activity (figure~\ref{fig:psd}).
  It would be biased and impractical to extrapolate these observations of jet breakdown location to other numerical studies~\citep{passerini13a, zmijanovic17a, fehn19a} as
  the LBM essentially solves the Boltzmann equation to recover NSE.
  Also, the parameters that have been of discussion in related studies~\citep{fehn19a, delorme13a} like Courant number cannot be directly related with LBM.
  It is however important to remark that numerical dissipation in LBM, even at
  the scales of grid spacing, and the numerical dispersive effects are smaller
  compared to other second-order accurate methods~\citep{marie}, which to a certain extent explains the consistent jet breakdown locations with increasing resolutions.

  In this study, we upsurged from NR to HR and XR directly without exploring a mid resolution range.
  It is noted that these are not a minimum resolution requirement for LBM simulations, and it is thus re-emphasized that this study was designed
  to explore LBM's suitability in performing fully resolved DNS, and thus resolutions, \emph{as high as possible}, within the confines of present computational paradigms were employed.
  It may be noted that the LB literature has grown considerably in the past and several ways to incorporate turbulence models within LB have been devised~\citep{succi08a}.
  Evaluations of these models using the FDA nozzle is left for future efforts.

  \subsubsection*{Onset of flow transition}
  If we focus on the FDA nozzle only, previous studies have found conflicting $Re_{crit}$, and why the numerically computed flow
  field did, or did not transition at $Re_{th}=2000$ in one particular study or the other is still unknown.
  It is fair to state that a flow in a perfectly symmetric setup as that of the FDA benchmark should not transition to turbulence in a simulation, and such an event must be a consequence of the numerics.
  Different numerical methods, parameters, stabilization techniques and resolutions may thus result in suppression or amplification of a turbulence triggering mechanism.
  Intense discussions have curtailed in the past about the role of resolutions
  to capture transitional phenomena~\citep{ventikos14a, delorme13a,
  zmijanovic17a, jain16a} while the physics, non-linear dynamics of a
  transitional flow, and its mechanobiological significance, if any, must be
  viewed with equal attention.
  A closer look at figure~\ref{fig:2khr} and the animations shows disturbance in the flow jet that emanates from the nozzle throat discussed above.
  These disturbances in the jet were also seen in the work of~\citet{fehn19a} despite the flow remained laminar in their simulations at this $Re_{th}$.
  Upon slight increase in $Re_{th}$ to $\sim 2400$ the jet did breakdown in their simulations to transition the flow. 
  It may be inferred that at $Re_{th}=2000$, the flow is on the verge of breakdown, which, depending on the numerics and inflow conditions
  used may or may not quantitatively breakdown.
  A complementary question is the circumstances that perturb the flow in the first place to trigger turbulence.
  It may be hypothesized that the perfect symmetry of the mesh in higher order methods suppresses the onset of transition but a conclusive statement on that cannot be made.
  Despite the fact that symmetry was ensured in the LB setup, there could have been artifacts from the boundaries that manifested as instabilities in the flow thereby triggering turbulence.
  \citet{white11a} demonstrated the inferior rotational invariance of the $D3Q19$ lattice and found $D3Q27$ superior whereas~\citet{dellar14a} advocated that multi time relaxation (MRT) model of the LBM recovers
  the rotational invariance.
  \citet{peng18a} also recently found both these types of lattice to yield accurate turbulent flow statistics.
  None of the study to the author's knowledge has investigated the role of lattice types in combination with the higher order wall boundary condition that was employed in this study~\citep{bouzidi13a}
  and also none has used these high resolutions.
  It may be inferred that a combination of MRT, higher order wall representation, and high resolutions must have overcome the minor deficiencies of the $D3Q19$ lattice but a detailed investigation of that is left for future efforts.
  It is also emphasized that the LBM is inherently a transient scheme, which might explain a closer match to the experiments.
  In a setup like the FDA nozzle it is clear that the sudden expansion resulted in adverse gradients of pressure, which, at a sufficiently high $Re_{th}$ 
  departed the flow from its laminar regime.

  {\color{black}

    It is finally remarked that the FDA nozzle is essentially a device to study
    blood flow and the non-Newtonian affects due to blood cells should be
    accounted for in future~\citep{saqr19a}.
    In LBM simulations non-Newtonian models that account for shear thinning
    behavior of fluid can be easily incorporated, and such a model is expected
    to \emph{delay} the onset of flow transition.
    On the other hand, however, \citet{tupin20a} have recently demonstrated a
    unique inverse energy cascade in blood flow and have found the turbulence
    of non-Kolmogorov type.
    The studies of transitional bioflows in future may study explicit transport
    of red blood cells, which requires incorporation of, and coupling with,
    other methods with LBM~\citep{sun}.
    The present work has prospects to serve as reference for future extensions
    and comparisons.
  }

\section{Conclusions and implications}
  %
  \begin{enumerate}
      \item The LBM is an adequate numerical technique for the DNS of
      biomedical flows in transitional regime, and can reproduce flow
      characteristics without much parameter tuning even at higher Reynolds
      numbers.
      The FDA benchmark, for transitional and turbulent flow regimes is
      suitable but not fully vigorous for a detailed quantitative comparison of
      CFD codes.
      The definition of the benchmark should have a reliable and quantifiable
      turbulence triggering mechanism, which may be incorporated into CFD
      models.

      \item The practice of \emph{quantitatively} and \emph{qualitatively}
      comparing experiments and simulations for a transitional flow is
      not entirely appropriate.
      Previously conducted experiments can only provide guidance on the setup
      of simulation and help in the analysis of results thereof.
      A comparison can only be performed by extensive joint efforts of
      experimentalists and computational researchers to \emph{tweak}
      experimental aspects like noise at inflow according to the
      simulations, or adjusting simulation aspects like initial conditions or
      inflow distortions.
      Previously reported discrepancies between experimental and computational
      results are non quantifiable and their source, while can be conjectured,
      it cannot be ascertained.

      \item A transitional flow is characterized by chaotic eddies and vortices
      with rapid annihilation and merger of vortices, and their interaction
      with the flow and each other, which results in distortions within the jet
      that emanates from the throat, continous restabilization and
      re-disruption of the jet at regular intervals.
      This results in shifts in the jet breakdown location, which are more
      pronounced at lower Reynolds numbers close to the $Re_{crit}$.

      \item Questions like \emph{jet breakdown location} have received major
      attention in the literature.
      The $Re_{crit}$, however, should be given equal attention and causes of
      discrepancies in its identification should be scrutinized to have better
      understanding of mechanobiological aspects like, for example, hemolysis
      or endothelial dysfunction.

      \item The LBM being a second order method requires relatively higher mesh
      density compared to other numerical methods but it can compute flows
      accurately, and in particular it can predict the onset of transition
      accurately in a relatively lesser time.
      On the other hand the method scales impeccably on massively parallel
      computer architectures thereby allowing simulations on complex geometries
      at any scale.
      Whether these facets of the LBM are assets or liabilities would depend on
      the perspective, the problem in hand and the research question itself.

      \item LBM based DNS approach would likely become \emph{impractical} for
      turbulent flows in complex geometries at Reynolds numbers higher than
      $6500$, and complex collision models or turbulence models should be
      explored in future.

  \end{enumerate}

\subsection*{acknowledgements}
    Compute resources on the \emph{SuperMUC-NG} were provided by the Leibniz
    Supercomputing Center (LRZ), Munich, \textsc{Germany}.
    The author is grateful for the able support of colleagues at LRZ.

\subsection*{Conflict of Interest Statement}
  The authors declare that they have no conflict of interest.
  No funding was received for this specific work.

\bibliographystyle{apalike2}       
\bibliography{references}%

\end{document}